\documentclass[12pt]{iopart}
\usepackage{amssymb}
\usepackage{setstack}
\usepackage{verbatim}
\usepackage{bm}
\usepackage{graphicx}

\begin{document}

\title[Multi-mode entanglement of $N$ harmonic oscillators]{Multi-mode entanglement of $N$ harmonic oscillators coupled to a non-Markovian reservoir}

\author{Gao-xiang Li$^{1}$, Li-hui Sun$^{1}$ and Zbigniew Ficek$^{2}$}
\eads{\mailto{gaox@phy.ccnu.edu.cn}}
\address{$^{1}$Department of Physics, Huazhong Normal University, Wuhan 430079, P. R. China}
\address{$^{2}$The National Centre for Mathematics and Physics, KACST, P.O. Box 6086, Riyadh 11442, Saudi Arabia}

\date{\today}

\begin{abstract}
Multi-mode entanglement is investigated in the system composed of $N$ coupled identical harmonic oscillators interacting with a common environment. We treat the problem very general by working with the Hamiltonian without the rotating-wave approximation and by considering the environment as a non-Markovian reservoir to the oscillators. We invoke an $N$-mode unitary transformation of the position and momentum operators and find that in the transformed basis the system is represented by a set of independent harmonic oscillators with only one of them coupled to the environment. Working in the Wigner representation of the density operator, we find that the covariance matrix has a block diagonal form that it can be expressed in terms of multiples of $3\times 3$ and $4\times 4$ matrices. This simple property allows to treat the problem to some extend analytically. We illustrate the advantage of working in the transformed basis on a simple example of three harmonic oscillators and find that the entanglement can persists for long times due to presence of constants of motion for the covariance matrix elements. We find that, in contrast to what one could expect, a strong damping of the oscillators leads to a better stationary entanglement than in the case of a weak damping.
\end{abstract}


\submitto{\JPB}

\maketitle


\section{Introduction}

Controlled dynamics and preservation of an initial entanglement encoded into a continuous variable system of harmonic oscillators coupled to a noisy environment are challenging problems in quantum information technologies~\cite{bp03,bl05}. The coupling induces decoherence phenomena, such as decay and dissipation that reduce and even can destroy the initial entanglement over a finite evolution time~\cite{dh98,w99,OpenQuantumSystems_book03}.  Dynamics of an open quantum system are usually studied in terms of the master equation of the reduced density operator whose structure depends on the nature of the environment to which the system is coupled. It has been noted that the dynamics crucially depend on whether the oscillators interact with a common or independent local environments. In the later case the interaction usually leads to a degradation of the entanglement whereas in the former, the environment can not only create decoherence, as it usually does, but may act as a source of coherence that not only preserves the initial entanglement but also creates an additional entanglement. A series of papers accounts these properties for the case of two coupled harmonic oscillators being in contact with a Markovian thermal reservoir and the work of Liu and Goan~\cite{lg07}, Maniscalco {\it et al.}~\cite{mo07} and H\"orhammer and B\"uttner~\cite{hb08} accounts for a non-Markovian thermal bosonic reservoirs. Detailed discussions and extensive reference lists devoted to the decoherence of two harmonic oscillators can be found in Refs.~\cite{b02,kl02,bf03,ok06,bl07,p02,sl04,dh04,pb04,bf06,az07,cb08,pr08}. Non-Markovian quantum dynamics of open systems has been discussed by others, notably by Breuer and Vacchini~\cite{bv08}, who provide the memory kernel treatment and illustrate it for various examples and applications.
In all these studies a general conclusion made is that entanglement dynamics depends on the form of the reservoir and the non-Markovian nature of the reservoir preserves entanglement over a longer time.

More important in the quantum technologies is the characterization and the study of dynamics of a large number of harmonic oscillators that are crucial for the study of quantum coherence, entanglement, fluctuations and dissipation of mesoscopic systems. The correct understanding of the mechanism responsible for entanglement evolution in the system is essential for designing $N$-atom systems for quantum information processing and quantum computation.
The key problem is to find the master equation for $N$ harmonic oscillators coupled to an environment that can be solved in a simple and effective way. How to treat such a composed system in the most effective way and how to understand its complicated dynamics are challenging questions that still have not been resolved.

In this paper, we pursue a research that especially addresses these questions. In the approach, we treat the problem very general, fully accounting the non-RWA dynamics and considering the environment as a non-Markovian reservoir to the oscillators. We introduce an $N$-mode unitary transformation of the position and momentum operators and find that in the transformed basis the system is represented by a set of independent harmonic oscillators with only one of them coupled to the environment. This fact makes the problem remarkably simple that the relaxation properties of $N$ harmonic oscillators follow the same pattern as a single harmonic oscillator.
This property also leads to two different time scales of the evolution of the system, a short time scale where the dynamics are strongly affected by the relaxation process, and a free of the relaxation long time scale. We distinguish those two time scales by working within the correlation matrix representation, also known as the covariance matrix.
We also consider squeezing of the position and momentum variances as a practical measure of a three-mode entanglement. We compare squeezing with the negativity~\cite{SeparabilityProperties_Loock03,SeparabilityProperties_Giedke01} and show that suitably transformed (rotated) quadrature components of the field modes exhibit squeezing whenever there is entanglement between the modes and vice versa.

\section{The model}\label{sc2}

We consider a system composed of $N$ mutually coupled identical harmonic oscillators of mass $M$ and frequency $\Omega$ that are simultaneously interacting with a common thermal bath environment (reservoir). The system is determined by the Hamiltonian, which in terms of the position $q_{i}$ and momentum $p_{i}$ operators  can be written~as
\begin{eqnarray}
H=H_{s}+H_{\varepsilon}+V_s +V ,\label{e1}
\end{eqnarray}
where
\begin{eqnarray}
H_{s}=\sum^{N}_{i=1}\left(\frac{p^{2}_{i}}{2M}\!+\!{\frac{1}{2}}{M}{\Omega^{2}} q^{2}_{i}\right)
\label{e2}
\end{eqnarray}
is the free Hamiltonian of the harmonic oscillators,
\begin{eqnarray}
H_{\varepsilon}=\sum_{n}\left(\frac{p^{2}_{n}}{2m_{n}}+{\frac{1}{2}}{m_{n}}
{\omega^{2}_{n}}{q^{2}_{n}}\right) \label{e3}
\end{eqnarray}
is the Hamiltonian of the common reservoir to which the oscillators are coupled,
\begin{eqnarray}
V_{s}=\lambda\sum^{N}_{i=1}\sum_{j>i}q_{i}q_{j} \label{e4}
\end{eqnarray}
is the interaction between the oscillators, and
\begin{eqnarray}
V=\sum_{n}\sum_{i=1}^{N}{\lambda_{n}}{q_{n}}q_{i} \label{e5}
\end{eqnarray}
is the interaction between the oscillators and the reservoir.

In equations~(\ref{e1})-(\ref{e5}),  the parameter $\lambda$ stands for
the coupling constant between the oscillators, and $\lambda_{n}$ is
the coupling strength of the oscillators to the reservoir. We model
the environment as an ensemble of harmonic oscillators of mass
$m_{n}$ and frequency $\omega_{n}$ that interact bilinearly through
their position operators $q_{n}$ with the oscillators.

The system of harmonic oscillators coupled to an environment is
usually described in terms of a reduced density operator
$\hat{\rho}$, which is obtained by tracing the density operator of
the total system over the reservoir operators. Instead of working in
the bare basis, $(q_{i},p_{i})$, we introduce an $N$-mode unitary
transformation of the systems' position operators
\begin{eqnarray}
\tilde{q}_k &=\sqrt{\frac{N-k}{N-k+1}}\left[q_k-\frac{1}{N-k}\sum^{N}_{j=k+1}q_j\right] ,\
 k =1,2,3,\ldots,N-1 ,\nonumber\\
\tilde{q}_N &= \frac{1}{\sqrt{N}}\sum^{N}_{k=1}q_k ,
\end{eqnarray}
and the same for the momentum operators. We note that the transformations involve anti-symmetrical $(\tilde{q}_{i},\tilde{p}_{i})$ and symmetrical $(\tilde{q}_{N},\tilde{p}_{N})$ combinations of the position and the momentum operators, a close analog of the symmetric and antisymmetric multi-atom Dicke states~\cite{dic,leh,ft02}.

In order to derive the master equation for the density operator $\hat{\rho}$ of the system, we use the standard method involving the Born approximation that corresponds to the second-order perturbative approach to the interaction between the oscillators and the environment, but we do not make the rotating-wave (RWA) and Markovian approximations. We find that in terms of the transformed operators the reduced density operator $\hat{\rho}$ satisfies the master equation
\begin{eqnarray}
\label{eq:masterequation}
\dot{\hat{\rho}}(t)&=& -\frac{i}{\hbar}\left[\tilde{H}_s+{\frac{1}{2}}M{\tilde{\Omega}^{2}_{N}(t)}{\tilde{q}_N^{2}},\hat{\rho}\right] - \frac{i}{\hbar}\gamma (t)\left[\tilde{q}_N,\{\tilde{p}_N,\hat{\rho}\}\right]\nonumber\\
&&-D(t)[\tilde{q}_N,[\tilde{q}_N,\hat{\rho}]]
-\frac{1}{\hbar}f(t)[\tilde{q}_N,[\tilde{p}_N,\hat{\rho}]] 
\end{eqnarray}

in which the Hamiltonian $\tilde{H}_{s}$ of the coupled oscillators
is of the form
\begin{eqnarray}
\tilde{H}_{s} = \sum^{N}_{i=1}\left(\frac{\tilde{p}^{2}_{i}}{2M} + {\frac{1}{2}}{M}{\Omega^{2}_{i}} \tilde{q}^{2}_{i}\right) ,
\end{eqnarray}
where
\begin{eqnarray}
\Omega_i &\equiv \Omega_{F} =  \sqrt{\Omega^2-\frac{\lambda}{M}} ,\  i=1,2,\ldots,N-1 ,\nonumber\\
\Omega_N &=\sqrt{\Omega^2+(N-1)\frac{\lambda}{M}} ,\label{e9}
\end{eqnarray}
are the effective frequencies of the oscillators. Note that the
frequency $\Omega_{N}$ of the oscillator coupled to the environment
differs from that of the remaining independent oscillators. It means
that the reservoir affects the evolution of only one of the
oscillators  leaving the remaining $N-1$ oscillators to evolve
freely in time.  The dynamics of the $N-$th oscillator that is
affected by the reservoir are determined by the following
time-dependent coefficients
\begin{eqnarray}
{\tilde{\Omega}^{2}_{N}(t)}&=-\frac{2}{M}\int^t_0{dt_1}\cos(\Omega_N t_1)\Pi(t_1) ,\label{e13}
\end{eqnarray}
represents a shift of the frequency of the oscillator due to the interaction with the environment. It
includes the frequency renormalization that leads to a finite Lamb
shift~\cite{Pathintegralapproach_Caldeira1983}.

The time dependent parameter 
\begin{eqnarray}
\gamma_N(t)&=\frac{1}{M\Omega_N}\int^t_0{dt_1}\sin(\Omega_N t_1)\Pi(t_1) \label{e14}
\end{eqnarray}
is the dissipation coefficient, and
\begin{eqnarray}
D_N(t)&=\frac{1}{\hbar}\int^t_0{dt_1}\cos(\Omega_N t_1)\nu(t_1) ,\\
f_N(t)&=-\frac{1}{M\Omega_N}\int^t_0{dt_1}\sin(\Omega_N t_1)\nu(t_1) ,
\end{eqnarray}
are diffusion coefficients.

The time dependent functions $\Pi(t)$ and $\nu(t)$
appear as the dissipation and noise kernels, respectively, and are given by
\begin{eqnarray}
\Pi(t) &= \frac{1}{2\hbar}\sum_{n}\lambda^2_n\langle[q_n(t),q_n(0)]\rangle
= \int^\infty_0{d\omega}J(\omega)\sin(\omega\,t) ,\\
\nu(t)&=\frac{1}{2\hbar}\sum_{n}\lambda^2_n\langle\{q_n(t),q_n(0)\}\rangle
=\int^\infty_0{d\omega}J(\omega)\cos(\omega\,t)[1+2\bar{N}(\omega)] ,
\end{eqnarray}
where $J(\omega)$ is the spectral density of the modes of the environment.
 For a Gaussian-type spectral density
\begin{eqnarray}
J(\omega)=\frac{2}{\pi}\gamma_0\omega M\left(\frac{\omega}{\Lambda}\right)^{n-1}
{\rm e}^{-\omega^2/\Lambda^2} ,
\end{eqnarray}
where $\Lambda$ is cut-off frequency that represents the highest
frequency in the environment,~$\gamma_0$ is proportional to the
coupling strength between the $N-$th oscillator  and the
environment, and~$n$ determines the type of the reservoir. For
$n=1$, the environment is called an Ohmic reservoir, for $n>1$ it is
called supra-Ohmic, whereas for $n<1$ it is called a sub-Ohmic
reservoir.

It should be stressed here that the derivation of the master equation holds under the Born approximation which takes the interaction between the oscillators and the reservoir only to the second order of the coupling strength $\lambda_{n}$. 

In the transformed basis, the Hamiltonian of the system exhibits
interesting properties. First of all, we observe that the system is
represented by a set of $N$ {\it independent} oscillators with {\it
only one} of them being coupled to the environment. The oscillator
effectively coupled to the environment is that one corresponding to
the symmetric combination of the position and momentum operators. In
addition, the effective frequency~$\Omega_{N}$ of the oscillator
coupled to the environment differs from that of the remaining
independent oscillators. The oscillators effectively decoupled from
the environment may be regarded as composing a relaxation-free
subspace. It should be stressed that the subspace is {\it not} a
decoherence-free subspace. We shall demonstrate that the subsystem
of the "relaxation free" oscillators still can evolve in time that
may lead to decoherence. We will recognize that only a part of the
subspace can be regarded as a decoherence-free subspace.

\section{Covariance matrix}\label{sc4}

We study dynamics of the system in terms of the Wigner
characteristic function, which for an $N$-mode Gaussian state can be
written in terms of a covariance matrix as~\cite{Atom_Photon_Cohen92}
\begin{eqnarray}
\chi(X)=\exp\left(-\frac{1}{2}\vec{X}V\vec{X}^T\right) ,
\end{eqnarray}
where $\vec{X} = {\rm col}(\tilde{q}_1,\tilde{p}_1,\tilde{q}_2,\tilde{p}_2,\ldots,\tilde{q}_N,\tilde{p}_N)$ is an $2N$ dimensional column vector of the transformed operators,
and $V$ is  the covariance matrix whose  elements are defined as
\begin{eqnarray}
V_{i,j} = {\rm Tr}\left(\{\Delta{X_i},\Delta{X_j}\}\hat{\rho}\right) ,\label{e24}
\end{eqnarray}
with 
\begin{eqnarray}
&&\Delta{X_i} = {X_i}-\langle{X_i}\rangle ,\nonumber\\
&&\{\Delta{X_i},\Delta{X_j}\} = \frac{1}{2}\left(\Delta{X_i}\Delta{X_j}+\Delta{X_j}
\Delta{X_i}\right) ,
\end{eqnarray}
and ${X}_{i}$ is the $i$th component of the vector $\vec{X}$.

The covariance matrix is composed of $4N^{2}$ elements. However, due
to the symmetrical property that $V_{ij}=V_{ji}$, it is enough to
find the diagonal elements and those off-diagonal elements with
$i<j$ to completely determine the matrix. Thus, the number of
elements that have to be found is equal to $N(2N+1)$. Technically,
it is done by using the definition (\ref{e24}) and the master
equation~(\ref{eq:masterequation}) form which one finds the
equations of motion for the covariance matrix elements that then are
solved for arbitrary initial conditions. However, the equations form
a set of coupled linear differential equations whose number is large
even for a small number of oscillators. Therefore, the dynamics of
coupled harmonic oscillators have usually been studied by employing
numerical methods.

We propose a different approach that illustrates the advantage of
working in the basis of the transformed position and momentum
operators. As we shall see, the approach allows to determine the
covariance matrix elements in an effectively easy way requiring to
solve separate sets of equations composed of only a small number of
coupled differential equations.

From equation~(\ref{e24}) and the master equation~(\ref{eq:masterequation}), we find a set of inhomogeneous differential equations for the covariance matrix elements, which can be written in a matrix form as
\begin{eqnarray}
\dot{\vec{V}}_{N}(t) = {\bf C}_{N}(t)\vec{V}_{N}(t) +\hbar \vec{F}_{N}(t),\label{e43}
\end{eqnarray}
where
\begin{eqnarray}
\vec{V}_{N}(t) ={\rm col}(V_{11},V_{12},V_{22},\ldots ,V_{2N-1,2N-1},V_{2N-1,2N},V_{2N,2N})
\end{eqnarray}
is a column vector of the covariance matrix elements,
\begin{eqnarray}
\vec{F}_{N}(t) ={\rm col}(0,0,0,\ldots,- f_{N}(t),2\hbar D_{N}(t))
\end{eqnarray}
is a column vector composed of the inhomogeneous time-dependent terms, and ${\bf C}_{N}(t)$ is an $N(2N+1)\times N(2N+1)$ block diagonal matrix of the time-dependent coefficients. The matrix ${\bf C}_{N}(t)$ is a direct sum of small size matrices
\begin{eqnarray}
{\bf C}_{N}(t) = \left[\displaystyle\bigoplus_{n=2}^N \left({\bf A}_{1}(0)\oplus {\bf A}_{4}^{(N-n)}(0)\oplus {\bf A}_{3}(t)\right)\right]\oplus{\bf A}_{2}(t) ,\label{e43a}
\end{eqnarray}
where
\begin{eqnarray}
{\bf A}_{1}(0) &= \left(\begin{array}{ccc}
0&2M^{-1}&0\\
-M\Omega^2_{F}&0&M^{-1}\\
0&-2M\Omega^2_{F}&0
\end{array}\right) ,\label{e27}
\end{eqnarray}

\begin{eqnarray}
{\bf A}_{2}(t) &= \left(\begin{array}{ccc}
0&2M^{-1}&0\\
-M\bar{\Omega}^{2}_{N}(t)&-\gamma(t)&M^{-1}\\
0&-2M\bar{\Omega}^{2}_{N}(t)&-2\gamma(t)
\end{array}\right) ,\label{e28}
\end{eqnarray}

\begin{eqnarray}
{\bf A}_3(t) &= \left(\begin{array}{cccc}
0&M^{-1}&M^{-1}&0\\
-M\bar{\Omega}^{2}_{N}(t)&-\gamma(t)&0&M^{-1}\\
-M\Omega^2_{F}&0&0&M^{-1}\\
0&-M\Omega^2_{F}&-M\bar{\Omega}^{2}_{N}(t)&-\gamma(t)
\end{array}\right) ,\label{e29}
\end{eqnarray}
and
\begin{eqnarray}
{\bf A}_4(0) &= \left(\begin{array}{cccc}
0&M^{-1}&M^{-1}&0\\
-M\Omega^2_F&0&0&M^{-1}\\
-M\Omega^2_F&0&0&M^{-1}\\
0&-M\Omega^2_F&-M\Omega^2_F&0
\end{array}\right) .\label{e29a}
\end{eqnarray}
with $\bar{\Omega}^{2}_{N}(t) = \Omega^2_N + \tilde{\Omega}^2_N(t)$ and
$\gamma(t)=2\gamma_{N}(t)$. The superscript $(N-n)$ in ${\bf A}_{4}^{(N-n)}(0)$ is understood as the number of the ${\bf A}_{4}(0)$ matrices appearing in the direct sum. Thus, for $N=2$, no matrix ${\bf A}_{4}(0)$ is involved in ${\bf C}_{N}(t)$, one matrix ${\bf A}_{4}(0)$ is involved for $N=3$, and so on. 

There are several interesting and important conclusions arising from equation~(\ref{e43a}). Firstly, the equations of motion group into
decoupled subsets of smaller sizes involving only three and four
equations.  In other words, the block diagonal matrix ${\bf
C}_{N}(t)$ is composed  of $3\times 3$ and $4\times 4$ matrices.
Secondly, the evolution of an $N>2$ system of harmonic oscillators
is determined by the same matrices as that determining the evolution
of  $N=2$ oscillators~\cite{b02,kl02,bf03,ok06,bl07,p02,sl04,dh04,pb04,bf06,cb08}.
Thirdly, the matrices ${\bf A}_{1}(0)$ and ${\bf A}_{4}(0)$ are
independent of time. This means that the time evolution of the
covariance matrix elements whose dynamics are determined by ${\bf
A}_{1}(0)$ and ${\bf A}_{4}(0)$ can be found in an exact analytical
form. Fourthly, the matrices ${\bf A}_{1}(0)$ and ${\bf A}_{4}(0)$
are independent of the relaxation coefficient $\gamma$. Thus, they
reflect features of the $N-1$ oscillators that are effectively
decoupled from the environment, and as such could be regarded as
determining a {\it relaxation-free subspace}. Finally, the  matrices
${\bf A}_{2}(t)$ and ${\bf A}_{3}(t)$ are explicitly dependent on
time through the relaxation terms $\gamma(t)$. Therefore, they
represent dynamics of the oscillator effectively coupled to the
environment. The explicit time dependence of the matrices
(\ref{e28}) and (\ref{e29}) results from the non-Markovian nature of
the environment, and the matrix becomes time independent in the case
of a Markovian situation. Thus, in the case for a Markovian
reservoir, all the covariance matrix elements can be found
analytically.

It should be noted here that the relaxation-free subspace cannot  be
regarded as a decoherence-free subspace because the covariance
matrix elements can undergo a periodic time evolution that may
result in a periodic decoherence. To explore this, we look into the
properties of the matrix ${\bf A}_{1}(0)$. It is easy to note that
the determinant of the matrix ${\bf A}_{1}(0)$ is equal to zero.
Mathematically, it means that among the three matrix elements
involved, $V_{11}, V_{12}$ and $V_{22}$, there is a linear
combination whose equation of motion is decoupled from the remaining
equations. It is easy to show that the linear combination
\begin{eqnarray}
V^{+}_{11} &=& M\Omega_{F}^{2}V_{11} +\frac{1}{M}V_{22} ,\nonumber\\
V^{-}_{11} &=& M\Omega_{F}^{2}V_{11} - \frac{1}{M}V_{22}
\end{eqnarray}
obeys $\dot{V}^{+}_{11} =0$ and the remaining elements form a set of
two coupled equations
\begin{eqnarray}
\dot{V}^{-}_{11} &= 4\Omega_{F}^{2}V_{12} ,\quad \dot{V}_{12} = -V^{-}_{11} ,\label{e31}
\end{eqnarray}
where $V^{-}_{11}= M\Omega_{F}^{2}V_{11} -(1/M)V_{22}$.

The property of $\dot{V}_{11}^{+}=0$ indicates that the linear
combination $V_{11}^{+}$ is a constant of motion, i.e.
$V_{11}^{+}(t) =V_{11}^{+}(0)$. In other words, $V_{11}^{+}(t)$ does
not change in time and retains its initial value for all times.
Physically, if initially the system was prepared in a state such
that $V_{11}^{+}(0)\neq 0$ and with the other elements of the
covariance matrix equal to zero, it would remain in that state for
all times. For example, if the initial state is an entangled state,
the initial entanglement of the system will remain constant in time.
Therefore, the subspace composed of the $V_{11}^{+}(t)$ element can
be regarded as a {\it decoherence-free subspace}.

The remaining matrix elements $V^{-}_{11}$ and $V_{12}$ can undergo
a temporal evolution.  Since there is no damping involved in the
equations of motion  (\ref{e31}), the solution would lead to the
matrix elements continuously oscillating in time. It is easy to
find that the solution of equation~(\ref{e31}) has a simple form
\begin{eqnarray}
V^{-}_{11}(t) &= V^{-}_{11}(0)\cos(2\Omega_{F}t) + 2\Omega_{F}V_{12}(0)\sin(2\Omega_{F}t) ,\nonumber\\
V_{12}(t) &=
V_{12}(0) \cos(2\Omega_{F}t) - \frac{V^{-}_{11}(0)}{2\Omega_{F}} \sin(2\Omega_{F}t)
,\label{e32a}
\end{eqnarray}
from which we see the matrix elements continuously oscillate in time
with frequency~$2\Omega_{F}$. This indicates that the system will
never reach a stationary time-independent state unless
$V^{-}_{11}(0)=V_{12}(0)=0$.  We stress that the continuous in time
oscillations are not related to the non-Markovian nature of the
reservoir as the matrix ${\bf A}_{1}(0)$ determines dynamics of the
oscillators that are not coupled to the reservoir.

It is also found that the determinant of the matrix ${\bf A}_{4}(0)$
is equal to zero. Thus, following the above analysis, we can find
that the set of the equations of motion for the covariance matrix
elements determined by the matrix ${\bf A}_{4}(0)$ can be reduced to
two constants of motion and two equations of motion with the same
coefficients as in~equation~(\ref{e31}).

On the basis of the above analysis, we may draw a conclusion that
the set of equations of motion for the covariance matrix elements
can be converted into $(N-1)^{2}$ constants of motion and a smaller
size set of coupled equations determined by a matrix
\begin{eqnarray}
{\bf C}_{N}^{\prime}(t) = {\bf A}_{5}^{\left(\frac{N(N-1)}{2}\right)}(0)\oplus
{\bf A}_{3}^{(N-1)}(t)\oplus{\bf A}_{2}(t) ,
\end{eqnarray}
where ${\bf A}_{5}(0)$ is a $2\times 2$ matrix composed of the
coefficients of the two coupled equations of motion~(\ref{e31}).

\section{Multi-mode entanglement and squeezing }\label{sc5}

We have already shown that due to the presence of the constants  of
motion in the evolution of the covariance matrix elements, the
dynamics of the system, even after a long time, may strongly depend
on the initial state. Since we are interested in the evolution of an
initial entangled state and it is well known that multi-mode
squeezed states are examples of entangled states, We have already shown that due to the presence of the constants of motion in the evolution of the covariance matrix elements, the dynamics of the system, even after a long time, may strongly depend on the initial state. Since we are interested in the evolution of an initial entangled state and it is well known that multi-mode squeezed states are examples of entangled states, we consider two experimentally realizable initial squeezed vacuum states with markedly  different squeezing behaviors. We also demonstrate that with the two specific initial states, the problem of treating the dynamics of $N$ harmonic oscillators simplifies to analysis of the properties of only those constants of motion and the matrices which involve only the diagonal elements of the covariance matrix.

In the first example, we consider the most familiar multi-mode
continuous variable Greenberger-Horne-Zeilinger (GHZ) entangled
state~\cite{l06,Multipartite_Entanglement_Loock00}
\begin{eqnarray}
|\psi_1\rangle=U_1\prod_{i=1}^{N}|0_{b_i}\rangle ,\label{e32}
\end{eqnarray}
with
\begin{eqnarray}
U_1\!=\!\exp\left\{-\frac{r}{6}\!\left[\!\sum_{i\neq j=1}^{N}\!\left(4b^\dag_i
b^\dag_j\!-\!(b^\dag_i)^2\right)\!-\!{\rm H.c.}\right]\right\} ,
\end{eqnarray}
where $r$ is the squeezing parameter and the ket $|0_{b_i}\rangle$
represents the state with zero photons in each of the~$N$ modes.
This GHZ state for $N=3$ has been realized experimentally by two
groups~\cite{Jing,Aoki}.

In the second example, we assume that the system is initially prepared in a pure non-symmetric multipartite squeezed state of the form
\begin{eqnarray}
|\psi_2\rangle=U_2\prod_{i=1}^{N}|0_{b_i}\rangle ,\label{e34}
\end{eqnarray}
where the squeezing transformation is of the form
\begin{eqnarray}
U_2 \!=\!\exp\left\{\!r_{0}\!\sum_{i=1}^{N-1}b^\dag_i
b^\dag_N\!+\!\frac{1}{2}r_{s}\!\sum_{i\neq j=1}^{N-1}b^\dag_i
b^\dag_j \!-\! {\rm H.c.}\!\right\} .
\end{eqnarray}
Here, each of the $N-1$ relaxation-free modes is correlated to a degree $r_{0}$ with the damped mode, whereas the relaxation-free modes are correlated between themselves to a degree $r_{s}$. Practical schemes for generation of such a state have recently been discussed~\cite{pfister}. For example, it  could be created by use of concurrent interactions in a second-order nonlinear medium placed inside an optical resonator, which might be realized experimentally in periodically poled KTiOPO$_4$~\cite{pfister1}. We will use this example to demonstrate the dependence of stationary entanglement on the amount of correlations initially encoded into the relaxation-free modes.

One manifestation of the squeezed properties of the states is entanglement between different modes. We examine this entanglement property shortly, but first we examine the manifestation of the squeezed correlations in the form of the covariance matrix.

The choice of the initial state (\ref{e32}) is a consequence of the diagonal form of the initial covariance matrix. In particular, the initial values of the covariance matrix elements of the two-mode case are
\begin{eqnarray}
V_{11}(0) =V_{44}(0)\!=\!\frac{1}{2}{\rm e}^{-2r} ,\,
V_{22}(0) = V_{33}(0)\!=\!\frac{1}{2}{\rm e}^{2r} ,\label{e36}
\end{eqnarray}
whereas for the three-mode case the initial elements are
\begin{eqnarray}
V_{11}(0)&=V_{33}(0)\!=\!V_{66}(0)\!=\!\frac{1}{2}{\rm e}^{-2r} ,\nonumber\\
V_{22}(0)&= V_{44}(0)\!=\!V_{55}(0)\!=\!\frac{1}{2}{\rm e}^{2r} ,
\end{eqnarray}

With the asymmetric squeezed state (\ref{e34}), the initial covariance matrix is not diagonal and has the following symmetric form
\begin{eqnarray}
{\bf{V}}(0) &= \left(\begin{array}{cccccc}
V_{11}(0)&0&V_{13}(0)&0&V_{15}(0)&0\\
0&V_{22}(0)&0&V_{24}(0)&0&V_{26}(0)\\
V_{13}(0)&0&V_{33}(0)&0&V_{35}(0)&0\\
0&V_{24}(0)&0&V_{44}(0)&0&V_{46}(0)\\
V_{15}(0)&0&V_{35}(0)&0&V_{55}(0)&0\\
0&V_{26}(0)&0&V_{46}(0)&0&V_{56}(0)
\end{array}\right) ,\label{eqlily}
\end{eqnarray}
where the explicit expressions for the non-zero matrix elements are given in the Appendix~A.

Before proceeding further with the analysis of the entangled and squeezing properties of the system, we return for a moment to the solutions for the covariance matrix elements. We have seen that with the states (\ref{e32}), the initial covariance matrix is diagonal. An immediate consequence of the diagonal form of the initial
covariance matrix is that for $t>0$,  the diagonal elements will be
different from zero and only those off-diagonal elements whose the
equations of motion were coupled to the equations of motion for the
diagonal elements. It is easy to see from equation~(\ref{e43}) that the
non-zero elements are determined by the matrices ${\bf A}_{1}(0)$
and ${\bf A}_{2}(t)$. Thus, dynamics of a system composed of $N$
harmonic oscillators can be readily determined from properties of
the two simple $3\times 3$ matrices.

For a multi-mode quantum state, if the variance of the quadrature
operator $\tilde{q}_{i}$ meets the inequality $\Delta
(\tilde{q}_{i})<1/2$, then we say the state exhibits ordinary
multi-mode squeezing. The minimum variance corresponds to the
optimal multi-mode squeezing. However, the ordinary squeezing is
produced by both  one and two-mode correlations, whereas
entanglement is solely related to the two-mode
correlations~\cite{dg00}. Thus, the ordinary squeezing does not
necessarily mean entanglement. We may distinguish between the
contributions of the one and two-mode correlations to the variances
and then determine the multi-mode squeezing by performing suitable
unitary transformations of the mode operators.

We illustrate this procedure for the case of three modes since the GHZ state~(\ref{e32}) for $N=3$ is an example of multipartite entangled state
whose entanglement is shared by more than two parties. Moreover, the
three-mode GHZ state has been realized
experimentally~\cite{Jing,Aoki} and also has been successfully
applied to demonstrate quantum
teleportation~\cite{Multipartite_Entanglement_Loock00,ya04} and
quantum dense coding~\cite{Jing}. We will demonstrate the
equivalence between the three-mode squeezing and the negativity
criterion for entanglement. The two-mode case, $N=2$, has been
extensively studied in the literature~\cite{peng-book}. First, we
make a local squeezing transformation on each of the modes, which
results in transformed annihilation operators of the
form~\cite{kc07}
\begin{eqnarray}
\tilde{a}_1&=& a_1{\rm e}^{i\theta}(u_{1}u_{2} - {\rm e}^{2i(\varphi-\theta)}v_{1}v_{2}) 
+a^\dag_1{\rm e}^{-i\theta}(u_{1}v_{2}-{\rm e}^{-2i(\varphi-\theta)}v_{1}u_{2}) ,\nonumber\\
\tilde{a}_2 &=& \tilde{a}_{3} = a_2{\rm e}^{i\theta}(u_{1}u_{2}
+{\rm e}^{2i(\varphi-\theta)}v_{1}v_{2}) 
-a^\dag_2{\rm e}^{-i\theta}(u_{1}v_{2}+{\rm e}^{-2i(\varphi-\theta)}v_{1}u_{2}) ,
\end{eqnarray}
where $u_{i}=\cosh(r_{i})$ and $v_{i}=\sinh(r_{i})\ (i=1,2)$ and the
transformation has been made with the squeezing parameter~$r_{1}$
and the phase angle $\varphi$ on the mode 1, and with $r_{2}$ and
the phase angle $\theta$ on the modes 2 and 3.

We use the Wigner characteristic function, which in terms of
the above specifically chosen transformation can be written in a
Gaussian form as
\begin{eqnarray}
\chi(\vec{\mu},t)=\exp\left\{-\frac{1}{2}\vec{\mu}{\bf{G}}\vec{\mu}^T\right\} ,
\end{eqnarray}
where $\vec{\mu} =(\tilde{y}_1,\tilde{x}_1,\tilde{y}_2,\tilde{x}_2,\tilde{y}_3,\tilde{x}_3)$ is a vector composed of the real $\tilde{y}_j$ and imaginary $\tilde{x}_j \ (j=1,2,3)$ parts of the phase-space variables corresponding to operator $\tilde{a}_j$, and~$\bf{G}$ is the correlation matrix of the form
\begin{eqnarray}
\bf{G}&= \left(\begin{array}{cccccc}
a&0&c&0&c&0\\
0&b&0&d&0&d\\
c&0&a&0&c&0\\
0&d&0&b&0&d\\
c&0&c&0&a&0\\
0&d&0&d&0&b
\end{array}\right) .
\end{eqnarray}
Note, the matrix $\bf G$ involves only four parameters that are
\begin{eqnarray}
a &= -2f_{3}{\rm e}^{2r_{2}} ,\quad b= -2f_{3}{\rm e}^{-2r_{2}} ,\nonumber\\
c &=2h_{1}{\rm e}^{2r_{2}} ,\quad d= 2h_{2}{\rm e}^{-2r_{2}} ,
\end{eqnarray}
where $h_{1} = (|f_1|-f_2)$ and $h_2 = -(|f_1|+f_2)$, with
\begin{eqnarray}
f_1&= m_2u_{1}^{2}+m_2^\ast {\rm e}^{4i\varphi}v_{1}^{2}
+2m_4{\rm e}^{2i\varphi}u_{1}v_{1} ,\nonumber\\
f_2 &= \left(m_2{\rm e}^{-2i\varphi}\!+\!m_2^\ast {\rm e}^{2i\varphi}\right)u_{1}v_{1}
+ m_4(1+2v_{1}^{2}) ,\nonumber\\
f_3 &= 4|m_1|u_{1}v_{1} +m_3(1+2v_{1}^{2}) ,
\end{eqnarray}
and $m_{i}$ are linear combinations of the covariance matrix elements $V^{\prime}_{ij}$ given in the bare basis
\begin{eqnarray}
m_1 &= \frac{1}{2}(V'_{11}-V'_{22}-2iV'_{12}) ,\nonumber\\
m_2 &=V'_{13}-V'_{24}-i(V'_{14}+V'_{23}) ,\nonumber\\
m_3 &=V'_{11}+V'_{22} ,\ m_4 =V'_{13}+V'_{24} .
\end{eqnarray}

The entangled nature of the three-mode squeezed states is clearly exhibited by the presence of the off-diagonal terms in the correlation matrix ${\bf G}$.

The squeezing parameters $r_{1}, r_{2}$ and the phase angles
$\varphi, \theta$ appearing in the transformation of the field
operators can be carefully chosen to match the form of the
correlation matrix~${\bf G}$ with the form of the covariance matrix
$V^{\prime}$ in the bare basis. In this way we can achieve the
equivalence between three-mode squeezing and entanglement. This can
be done by choosing the squeezing parameters~as
\begin{eqnarray}
 {\rm e}^{2r_1} = \left(\frac{m_3-2|m_1|}{m_3+2|m_1|}\right)^{\!\frac{1}{2}}, \quad {\rm e}^{2r_2}
=\left(\frac{|h_2| +f_3}{|h_1| +f_3}\right)^{\!\frac{1}{2}} ,
\end{eqnarray}
with $ m_1 =|m_1|\exp(2i\varphi)$ and $f_1=|f_1|\exp(2i\theta)$.

Having available the time dependent solutions for the covariance
matrix  elements, we then can easily find the characteristic
function that allows us to compute variances of the position
operators and momentum operators
\begin{eqnarray}
\tilde{X}_{k}&=\sqrt{\frac{3-k}{2(4-k)}}\left[\!\left(\tilde{a}_k\!-\!\frac{1}{3-k}\sum^{3}_{j=k+1}\!\tilde{a}_j\right)\!+ {\rm H.c.}\right],\nonumber\\
\tilde{Y}_{k}&=-i\sqrt{\frac{3-k}{2(4-k)}}\left[\!\left(\tilde{a}_k\!-\!\frac{1}{3-k}\sum^{3}_{j=k+1}\!\tilde{a}_j\right)\!- {\rm H.c.}\right] ,
\end{eqnarray}
for $k=1,2$, and
\begin{eqnarray}
\tilde{X}_3 =\sqrt{\frac{1}{6}}\sum\limits_{j=1}^3\left(\tilde{a}_j+\tilde{a}_{j}^{\dagger}\right) ,\, \tilde{Y}_3=-i\sqrt{\frac{1}{6}}\sum\limits_{j=1}^3\left(\tilde{a}_j-\tilde{a}_{j}^{\dagger}\right)  . \label{eee}
\end{eqnarray}
The variances are involved in the criterion for multi-mode squeezing
that fluctuations of the correlations between three modes are
squeezed if and only if the sum of the variances $\langle(\Delta
\tilde{X}_{i})^2\rangle$ and $\langle(\Delta
\tilde{Y}_{j})^2\rangle$ with $i\neq j$ satisfies the following
inequality~\cite{SeparabilityProperties_Loock03}
\begin{eqnarray}
\langle(\Delta \tilde{X}_{i})^2\rangle+\langle(\Delta \tilde{Y}_{j})^2\rangle < 1 ,\  i,j =1,2,3 .\label{e56}
\end{eqnarray}
Among the permutations of the variances involved on the left-hand side of equation~(\ref{e56}),  there might be more than one satisfying inequality condition for multi-mode squeezing. In this case, we choose the  combination that reflects the largest squeezing.

To quantify entanglement between the modes, we adopt the negative partial transpose criterion that is known as the necessary and sufficient condition for entanglement of two- and three-mode Gaussian states.  We will compare the criterion with the squeezing criterion to quantify squeezing as an alternative necessary and sufficient condition for entanglement. The advantage of the squeezing criterion over the negativity is that the former can be directly measured in experiments whereas the later can be inferred from the reconstruction of the density matrix of the system.

The partial transpose criteria are based on the non-positive partial transpose of a
matrix~\cite{SeparabilityCriterion_Simon00,dg00}
\begin{eqnarray}
\Gamma_jV^\prime(t)\Gamma_j+\frac{1}{2}i\sigma  ,\ j=1,2,3,\ldots \label{e42}
\end{eqnarray}
where $\Gamma_j$ is the partial transpose matrix with the transposition made on the $j$th mode block and $\sigma$ is a block diagonal symplectic matrix.
It has been shown that multi-mode Gaussian states are not completely separated when for all $j$ there are negative eigenvalues of the matrix (\ref{e42}). The eigenvalues can be degenerated or non-degenerated. However, for a system of identical oscillators the covariance matrix $V^\prime(t)$ is permutational symmetric, so all the negative eigenvalues are degenerated. We denote them by a parameter $\eta_-$ and call it as the negativity criterion for entanglement.

\section{Temporal evolution of squeezing and entanglement}\label{sc6}

We now perform numerical analysis of time evolution of multi-mode
squeezing and entanglement in a system of two and three mutually
interacting harmonic oscillators simultaneously coupled to an
environment. We will illustrate the advantage of working in the
transformed basis to obtain a simple interpretation of the results.
In particular, to understand short time non-Markovian dynamics of
entanglement and to provide conditions for optimal and stable long
time entanglement. In addition, we compare the time evolutions of
the variances and the negativity to find if the condition for
three-mode squeezing could be used as the necessary and sufficient
condition for three mode entanglement.  In all cases considered
here, we assume that  the oscillators interact
with an Ohmic reservoir $(n=1)$ of temperature
$k_{B}T=10\hbar\Omega$ with the Boltzmann distribution of photons
characterized by the mean occupation number $\bar{N}(\Omega) =9.5083$.
\begin{figure}[h]
 \begin{center}
   \begin{tabular}{c}
   \includegraphics[height=10cm,width=0.8\columnwidth]{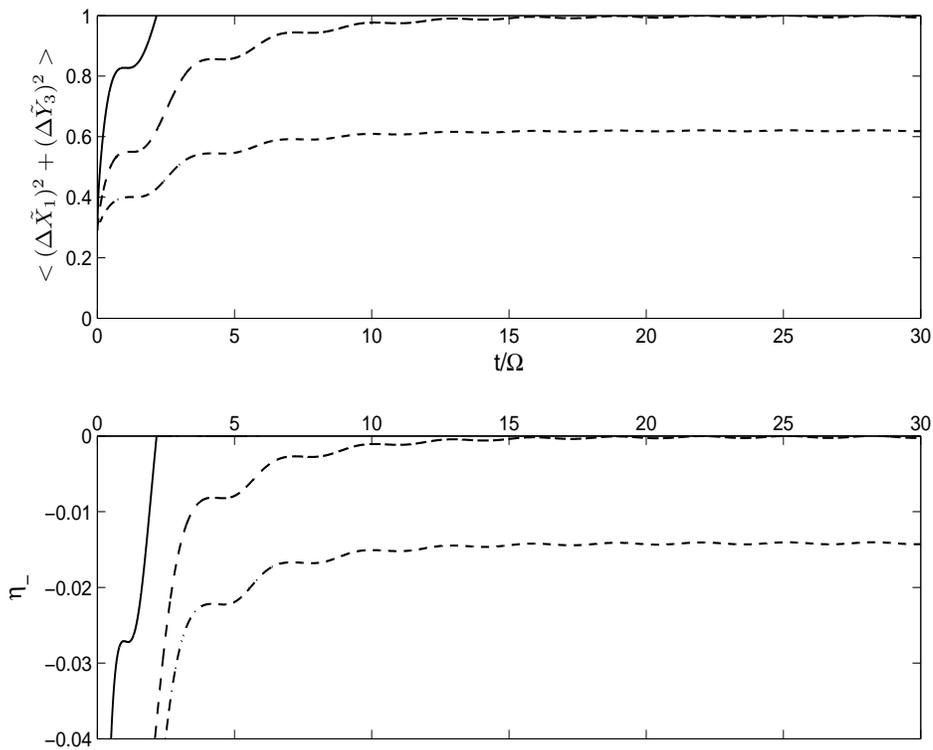}
   \end{tabular}
   \end{center}
\caption{Time evolution of the negativity $\eta_-$ and the combined variance $\langle(\Delta \tilde{X}_{1})^2\rangle+\langle(\Delta \tilde{Y}_{3})^2\rangle $ for $\gamma_0=0.05, \Lambda=100, n=1, \lambda=0$ and different $r$: $r=1.0$ (solid line), $r=1.498$ (dashed line), $r=2.0$ (dashed-dotted line). The system was initially in the state $|\psi_1\rangle$.}
\label{lifig2}
\end{figure}

We first consider the case of mutually independent oscillators  with
$\lambda=0$, but interacting with the environment.
 Figure~\ref{lifig2} shows the negativity and variances as a function
of time for the initial symmetric squeezed state $|\psi_{1}\rangle$
with different degree of squeezing $r$. First of all, we note that
at times where squeezing occurs there is entanglement, and vice
versa, at times where entanglement occurs, there is squeezing. In
addition, we see a threshold value for the degree of squeezing $r$
at which a continuous in time entanglement occurs. The threshold
that corresponds to entanglement undergoing the phenomenon of sudden
death,  occurs at $r=1.498$. It is interesting to note that the same
threshold value for $r$ has been predicted for the two-mode
case~\cite{pb04}.

The presence of the threshold value for $r$ at which continuous in
time entanglement occurs has a simple interpretation in terms of the
covariant matrix elements. Consider the threshold in the long time
limit in which we may consider the evolution under the Markov
approximation, but retaining the non-RWA terms. Under this
approximation, we can put $\gamma(t)\rightarrow \gamma_{0}$ which
then allows us to obtain a simple analytical solution for the
threshold condition for entanglement.

It is easy to show that the threshold for two mode entanglement occurs at
\begin{eqnarray}
V_{11}(t)V_{44}(t) = 1 /4,\label{e60}
\end{eqnarray}
so that the two modes are entangled when $V_{11}(t)V_{44}(t) <1/4$,
otherwise are separable. Note that the covariance matrix element $V_{11}(t)$ is associated with the relaxation free modes whereas the element $V_{44}(t)$ is associated with the mode that is coupled to the reservoir and thus undergoes the damping process. Under the Markov approximation, we find from equations~(\ref{e43})-(\ref{e29})  that in the long time limit of $t\gg \gamma^{-1}$, the element $V_{44}(t)$ reaches the stationary value equal to the level of the thermal fluctuations
\begin{eqnarray}
V_{44}(t)\rightarrow 2{\bar N}+1 ,\label{e61}
\end{eqnarray}
whereas $V_{11}(t)$ retains its time dependent behavior which depends on the initial values
\begin{eqnarray}
V_{11}(t) &= V_{11}(0)\cos^{2}\Omega_{F}t
+\frac{V_{22}(0)}{M^{2}}\left(\frac{\sin\Omega_{F}t}{\Omega_{F}}\right)^{2}
  .\label{e62}
\end{eqnarray}
We point out that the dependence on the initial values of the long
time behavior of $V_{11}(t)$ is due to the presence of the constant
of motion $V_{11}^{+}$.

Averaging equation~(\ref{e62}) over a long period of oscillations, the threshold condition~(\ref{e60}) simplifies to
\begin{eqnarray}
2V_{11}(0)\left(2{\bar N}+1\right) = 1.\label{e63}
\end{eqnarray}
We see that the threshold behavior of entanglement depends on the initial value of the covariance matrix element $V_{11}(0)$. In other words, the entanglement behavior can be controlled by the suitable choosing of the initial state. For example, with the initial state (\ref{e32}), we find from equations~(\ref{e63}) and (\ref{e36}) that continuous entanglement occurs for the degree of squeezing
\begin{eqnarray}
r = \frac{1}{2}\ln\left(2{\bar N}+1\right) .\label{e64}
\end{eqnarray}
With the parameter value $k_{B}T=10\hbar\Omega$, we find that the threshold value for $r$ equals to $1.498$ that is the same found numerically in figure~\ref{lifig2}. We should point out here that the same condition for the threshold value of $r$ has been found under the RWA approximation~\cite{pb04}. Thus, we may conclude that the threshold value for continuous entanglement is not sensitive to the RWA approximation.

\begin{figure}
   \begin{center}
   \begin{tabular}{c}
   \includegraphics[height=10cm,width=0.8\columnwidth]{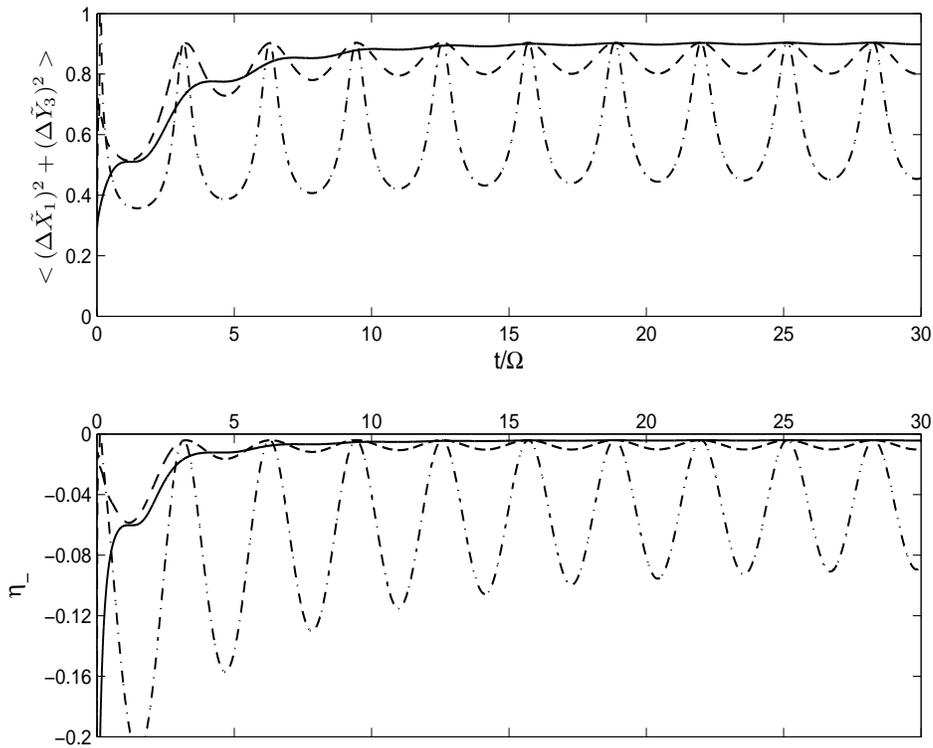}
   \end{tabular}
   \end{center}
   \caption{Time evolution of the negativity $\eta_-$ and combined variance $\langle(\Delta \tilde{X}_{1})^2\rangle+\langle(\Delta \tilde{Y}_{3})^2\rangle $  for $\Lambda=100, n=1,\lambda=0, r=1.6$ and different values of the relaxation rate $\gamma_{0}$: $\gamma_{0}=0.05$ (solid line), $\gamma_{0}=1.0$ (dashed line), $\gamma_{0}=5.0$ (dashed-dotted line). The system was initially in the state $|\psi_1\rangle$.}
   \label{lifig3}
   \end{figure}

We now proceed to discuss the dependence of the long time entanglement on the relaxation rate $\gamma_{0}$. An example of this feature is shown in figure~\ref{lifig3}. It is interesting to note that under the relaxation the entanglement oscillates in time and the amplitude of the oscillations increases with increasing $\gamma_{0}$ leading to a better entanglement when the oscillators are strongly damped. It is a surprising result as one could expect that entanglement should decrease with increasing $\gamma_{0}$. Again, a straightforward interpretation of this effect can be gained from a qualitative inspection of the properties of the transformed covariance matrix.

It is easy to see from equations~(\ref{e27}) and (\ref{e28}) that in the
limit of  vanishing damping, $\gamma(t)\rightarrow 0$ and $\lambda
=0$, the matrix ${\bf A}_{2}(t)$ reduces to ${\bf A}_{1}(0)$. One
could argue that in this limit the covariance matrix elements
determined by the matrix ${\bf A}_{2}(t)$ coincidence with the
elements determined by the matrix~${\bf A}_{1}(0)$. Of course, their
time evolution is determined by the same equations, but there is a
subtle difference in their initial values. For example, the initial
values of the elements whose evolution is determined by the matrix
${\bf A}_{1}(0)$ are
\begin{eqnarray}
V^{\pm}_{11}(0) &= \frac{1}{2}\left(M\Omega_{F}^{2}{\rm e}^{-2r} \pm \frac{1}{M}{\rm e}^{2r}\right) ,
\end{eqnarray}
whereas that one determined by the matrix $A_{2}$ are
\begin{eqnarray}
V^{\pm}_{55}(0) &= \frac{1}{2}\left(M\Omega_{F}^{2}{\rm e}^{2r} \pm \frac{1}{M}{\rm e}^{-2r}\right) .
\end{eqnarray}
The initial elements are significantly different that what appears as a squeezed component in $V^{\pm}_{11}(0)$, the counterpart in $V^{\pm}_{55}(0)$ appears as an anti-squeezed component. This is a crucial difference that has a significant effect on the evolution of an entanglement. These two contributions cancel each other that results in no oscillations in the entanglement evolution when $\gamma_{0}\ll 1$. On the other hand, for a large~$\gamma_{0}$ the covariance matrix elements determined by ${\bf A}_{2}(t)$ are rapidly damped to their stationary values leaving the elements determined by ${\bf A}_{1}(0)$ continuously oscillating in time. These oscillations lead to the continuous oscillation of the entanglement seen in figure~\ref{lifig3}.

One can interpret these results in terms of collective symmetric and antisymmetric states of an $N$-atom Dicke model~\cite{dic,leh,ft02}. The symmetric and antisymmetric states correspond to the atomic dipole moments oscillating in-phase and out-of-phase, respectively. The most interesting is that in the case of the atoms coupled to a common reservoir, the antisymmetric states do not decay, whereas the symmetric states decay with an enhanced rate $N\gamma$, where $\gamma$ is the single atom decay rate. Hence, in the absence of the damping, oscillations induced by the symmetric and antisymmetric states cancel each other as they occur with opposite phases. When damping is included, the oscillations induced by the symmetric states are damped in time whereas the oscillations induced by the antisymmetric states remain unaffected. The oscillations induced by the symmetric states die out on the time scale of $t\sim 1/(N\gamma)$ leaving the oscillations induced by the antisymmetric states unaffected. 
\begin{figure}
   \begin{center}
   \includegraphics[height=10cm,width=0.8\columnwidth]{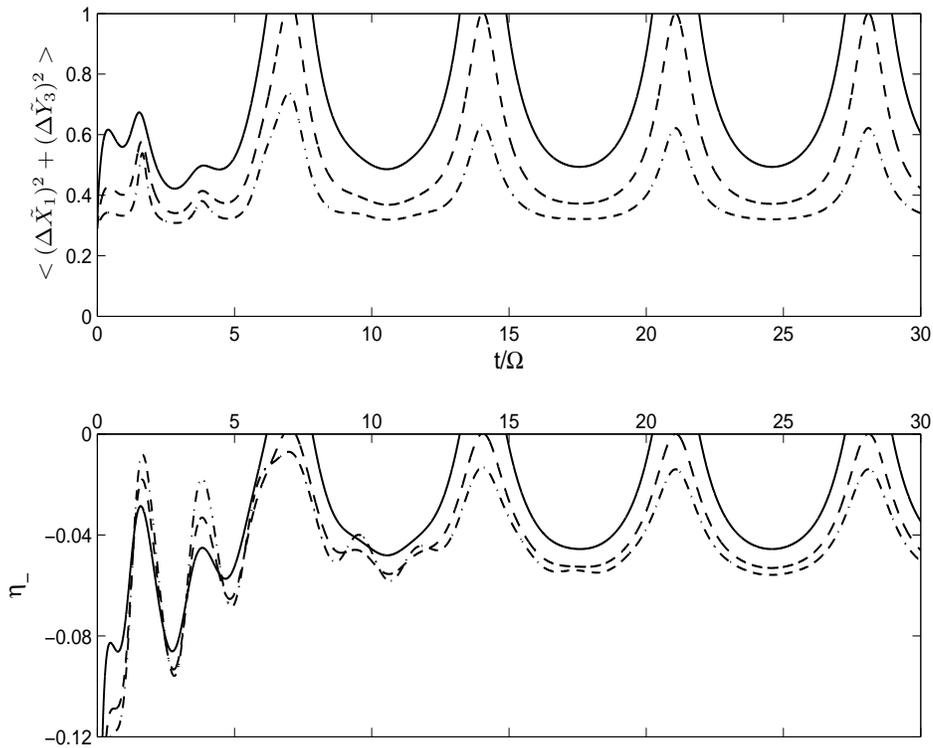}
   \end{center}
   \caption[nsa]{Time evolution of the negativity $\eta_-$ and the combined variance $\langle(\Delta \tilde{X}_{1})^2\rangle+\langle(\Delta \tilde{Y}_{3})^2\rangle  $  for $\gamma_0=0.05, \Lambda=100, n=1, \lambda=0.8$ and different $r$: $r=1.0$ (solid line), $r=1.498$ (dashed line), $r=2.0$ (dashed-dotted line). The system was initially in the state $|\psi_1\rangle$.}
\label{lifig4} 
\end{figure}

Figure~\ref{lifig4} shows the evolution of entanglement and
squeezing when  the oscillators are coupled to each other. In this
case there is no continuous stationary entanglement. Thus, the
interaction between the oscillators has a destructive effect on the
stationary entanglement. However, for a large squeezing,
entanglement re-appears in some discrete periods of time, exhibiting
periodic sudden death and revival of entanglement. In other words,
the threshold behavior of entanglement is a periodic function of
time. As before, this feature has a simple interpretation in terms
of the covariance matrix elements. According to equation (\ref{e60}), for
a given temperature the threshold value for entanglement depends on
the covariance matrix element $V_{11}(t)$ which, on the other hand,
depends on~$\lambda$ through the frequency parameter~$\Omega_{F}$.
We see from equation (\ref{e9}) that~$\Omega_{F}$ decreases with
increasing~$\lambda$. Thus, according to equation (\ref{e62}) for
interacting oscillators the matrix element $V_{11}(t)$ oscillates
slowly in time. The averaging over the oscillations is not justified
and thus the threshold condition for entanglement is the oscillating
function of time even in a long time regime.

   \begin{figure}
   \begin{center}
   \begin{tabular}{c}
   \includegraphics[height=10cm,width=0.8\columnwidth]{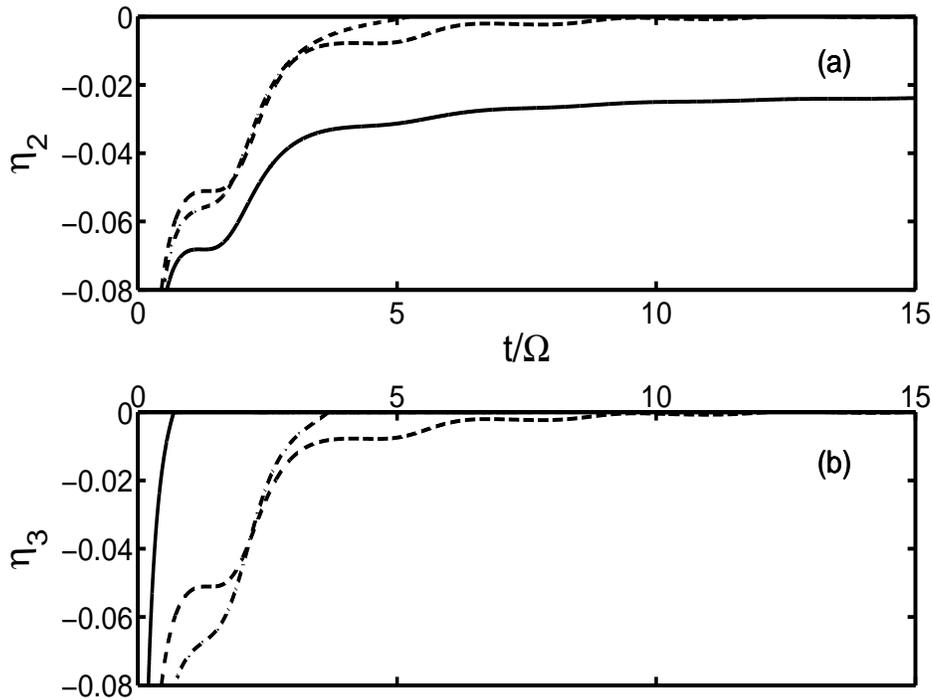}
   \end{tabular}
   \end{center}
   \caption[nsa]
   { \label{lifig5} Time evolution of the negativity (a) $\eta_2$ and (b) $\eta_3$ for the initial asymmetric state $|\psi_{2}\rangle$ with $\gamma_0=0.05, \Lambda=100$, $\lambda=0, r_{s}=1.489$ and different $r_{0}$: $r_{0}=1.0$ (solid line), $r_{0}=1.489$ (dashed line), $r_{0}=2.0$ (dashed-dotted line).}
\end{figure}

Finally, in figure~\ref{lifig5} we illustrate the evolution of entanglement for two different cases of the initial asymmetric state $|\psi_2\rangle$. As we have shown in section~\ref{sc4}, more constants of motion are then involved than in the symmetric case which, on the other hand, may lead to a better stationary entanglement. In the first case, we plot the negativity $\eta_{2}$ which describes entanglement between the mode $2$ and the pair $1\leftrightarrow 3$. We see from figure~\ref{lifig5}(a) that the stationary entanglement appears only when $r_{0}<r_{s}$. Otherwise, the initial entanglement rapidly decays to zero and disappears after a finite time.  Again, this feature can be easily explained in terms of the transformed oscillators. When $r_{0}<r_{s}$, the pair of modes 1 and 2 that is decoupled from the environment is more strongly correlated than the pairs $1\leftrightarrow 3$ and $2\leftrightarrow 3$, which involve the mode coupled to the environment. This preserves the entanglement in the system. In the opposite case of $r_{0}>r_{s}$, a large entanglement is initially encoded into the pairs that are damped due to the coupling to the environment. This results in the loss of the correlations and entanglement. Quite different properties exhibits entanglement between the mode $3$, which is coupled to the environment, and the remaining pair $1\leftrightarrow 2$. In this case, illustrated in figure~\ref{lifig5}(b) there is no stationary entanglement. This can be interpreted as the result of the coupling of the mode $3$ to the reservoir that leads to the continuous dissipation of the initial correlations $r_{0}$.

\section{Conclusion}\label{sc7}

We have analyzed dynamics of a set of $N$ harmonic oscillators coupled to a non-Markovian reservoir in terms of the covariance matrix. By performing a suitable transformation of the position and momentum operators of the system oscillators, we have shown that the set of  coupled differential equations for the covariance matrix elements splits into decoupled subsets of smaller sizes involving only three and four equations.  In other words, our analysis clearly show that the dynamics of $N$ oscillators can be  completely determined by properties of $4\times 4$ and $3\times 3$ matrices. The approach proposed here could be particularly useful in applications to macroscopic systems composed of a large number of oscillators for which numerical analysis are technically too complicated or impossible to perform.

The approach has been applied to the case of three coupled harmonic oscillators interacting with a non-Markovian reservoir. A general feature of the entanglement evolution is that it exhibits two characteristic time scales, a shot time regime where an initial entanglement is rapidly damped and a long time regime where the entanglement undergoes continuous undamped oscillations.
Depending on the initial amount of entanglement encoded into the system, it can be preserved for all times or may periodically disappear and reappear that the entanglement may undergo the sudden death and revival phenomena. We have also found that in contrast to what one could expect, a stronger damping of the oscillators leads to a better stationary entanglement than in the case of a weak damping.
Finally, we point out that the three-mode entanglement can be observed experimentally, simply by detecting quadrature squeezing of the field modes.

\section*{Acknowledgments}

We acknowledge financial support from the National Natural Science
Foundation of China (Grant No. 60878004), the Ministry of Education
under project  SRFDP (Grant No. 200805110002), the National Basic
Research Project of China (Grant No. 2005 CB724508).

\appendix

\section{Initial values of the covariance matrix}

In this appendix, we list the non-zero elements of the initial covariance matrix for the case of the asymmetric initial state~(\ref{e34}). The diagonal elements are of the form
\begin{eqnarray}
V_{11}(0)&=\frac{{\rm e}^{-2r_s}}{24}
\left[9 + {\rm e}^{3r_s}\left(3\cosh\bar{r}-q\sinh\bar{r}\right)\right] ,\nonumber\\
V_{22}(0)&=\frac{{\rm e}^{2r_s}}{24}
\left[9 + {\rm e}^{-3r_s}\left(3\cosh\bar{r}+q\sinh\bar{r}\right)\right] ,\nonumber\\
V_{33}(0)&=\frac{{\rm e}^{-2r_s}}{8}
\left[1 + {\rm e}^{3r_s}\left(3\cosh\bar{r}-q\sinh\bar{r}\right)\right] ,\nonumber\\
V_{44}(0)&=\frac{{\rm e}^{2r_s}}{8}
\left[1 + {\rm e}^{-3r_s}\left(3\cosh\bar{r}+q\sinh\bar{r}\right)\right] ,\nonumber\\
V_{55}(0)&=\frac{{\rm e}^{r_s}}{6}\left(3\cosh\bar{r}+q\sinh\bar{r}\right) ,\nonumber\\
V_{66}(0)&=\frac{{\rm e}^{-r_s}}{6}\left(3\cosh\bar{r}-q\sinh\bar{r}\right) ,
\end{eqnarray}
whereas the off-diagonal terms are
\begin{eqnarray}
V_{13}(0)&=\frac{-{\rm e}^{-2r_s}}{8\sqrt{3}}
\left[3 - {\rm e}^{3r_s}\left(3\cosh\bar{r}-q\sinh\bar{r}\right)\right] ,\nonumber\\
V_{24}(0)&=\frac{-{\rm e}^{2r_s}}{8\sqrt{3}}
\left[3 - {\rm e}^{-3r_s}\left(3\cosh\bar{r}+q\sinh\bar{r}\right)\right] ,\nonumber\\
V_{35}(0)&=\sqrt{3}V_{25}(0)=-\frac{{\rm e}^{r_s}(r_{0}-r_s)\sinh\bar{r}}{\sqrt{6}\bar{r}} ,\nonumber\\
V_{46}(0)&= \sqrt{3}V_{26}(0)=\frac{{\rm e}^{-r_s}(r_{0}-r_s)\sinh\bar{r}}{\sqrt{6}\bar{r}} ,
\end{eqnarray}
where $\bar{r}=\sqrt{8r_{0}^{2}+r_s^{2}}$ and $q=(8r_{0}+r_{s})/\bar{r}$.

\end{document}